\documentclass[11pt]{article}

\pdfoutput=1  %arxiv 

\usepackage[margin=0.97in,top=.99in,bottom=1.1in, textheight=22.51cm,textwidth=16.51cm, heightrounded,columnsep=0.81cm]{geometry}

\usepackage{cite,url}

\usepackage[usenames,dvipsnames]{color}
\usepackage{graphicx,wrapfig,subfigure}
\usepackage{flushend}

\usepackage[cmex10]{amsmath}
\usepackage{amsfonts,amssymb,mathrsfs}   %breqn
\usepackage[colorlinks,bookmarksopen,bookmarksnumbered,citecolor=red,urlcolor=red]{hyperref}
\usepackage{soul}

\newlength{\noteWidth}
\long\def\notes#1{\ifinner
           {\footnotesize #1}
           \else
           \marginpar{\parbox[t]{\noteWidth}{\raggedright\footnotesize #1}}
       \fi\typeout{#1}}

\def\notes#1{\typeout{read notes: #1}}  %uncomment for final version

\def\spm#1{\notes{SPM:  #1}}

\newcommand*{\qed}{\nobreak\hfill\ensuremath{\square}}

\newcounter{rmnum}

\newenvironment{romannum}{\begin{list}{{\upshape (\roman{rmnum})}}{\usecounter{rmnum}
\setlength{\leftmargin}{8pt}
\setlength{\rightmargin}{8pt}
\setlength{\itemsep}{2pt}
\setlength{\itemindent}{-1pt}
}}{\end{list}}
 
\newcounter{anum}

\def\tily{\tilde{y}}

\def\util{\mathchoice{\mbox{\small$\cal U$}}%
{\mbox{\small$\cal U$}}%
{\mbox{$\scriptstyle\cal U$}}%
{\mbox{$\scriptscriptstyle\cal U$}}}

\def\bfgamma{\bfmath{\gamma}}

\def\Spx{\textsf{S}}
\def\Ebox#1#2{%
\begin{center}
\includegraphics[width= #1\hsize]{#2} %\epsfxsize=\hsize \epsfbox{#2}}
\end{center}}

\def\Fig#1{Fig.~\ref{#1}}

\def\ind{\field{I}}

\def\Re{\field{R}}

\def\piload{\Gamma}
\def\Health{{\cal L}}
\def\health{\ell}
\def\psd{\text{S}}
\def\epsy{\varepsilon}

\def\bfmath#1{{\mathchoice{\mbox{\boldmath$#1$}}%
{\mbox{\boldmath$#1$}}%
{\mbox{\boldmath$\scriptstyle#1$}}%
{\mbox{\boldmath$\scriptscriptstyle#1$}}}}

\def\bfme{\bfmath{e}}

\def\bfmr{\bfmath{r}}

\def\bfmy{\bfmath{y}}

\def\bfmw{\bfmath{w}}

\def\bfmD{\bfmath{D}}

\def\bfmL{\bfmath{L}}

\def\bfmR{\bfmath{R}}

\def\bfmX{\bfmath{X}}

\def\bfmY{\bfmath{Y}}

\def\bfmhhaY{\bfmath{\hhaY}} %\widehat{\widehat{Y}}}}
\def\bfmhhaY{\hbox to 0pt{$\widehat{\bfmY}$\hss}\widehat{\phantom{\raise 1.25pt\hbox{$\bfmY$}}}}

\def\bfzeta{\bfmath{\zeta}}

\newtheorem{theorem}{Theorem}[section]

\newtheorem{proposition}[theorem]{Proposition}

\def\Proposition#1{Proposition~\ref{#1}}
\def\Theorem#1{Theorem~\ref{#1}}

\def\Section#1{Section~\ref{#1}}

\def\state{{\sf X}}

\newcommand{\field}[1]{\mathbb{#1}}

\def\Co{\field{C}}

\def\Re{\field{R}}

\def\intgr{\field{Z}}

\def\nat{\field{Z}_+}

\def\bary{{\overline {y}}}

\def\Prob{{\sf P}}

\def\Expect{{\sf E}}

\def\eqdef{\mathbin{:=}}

\def\transpose{{\hbox{\it\tiny T}}}

\def\psdder{\text{S}^{\text{\tiny(2)}}}

\def\clE{{\cal E}}

\def\clI{{\cal I}}

\def\clV{{\cal V}}

\title{Estimation
and Control of 
\\
Quality of Service  
 in Demand Dispatch}
\author{Yue Chen, 
Ana Bu\v{s}i\'c, and Sean Meyn% <-this % stops a space
\thanks{This research is supported by the NSF grants CPS-0931416 and CPS-1259040, the French National Research Agency grant ANR-12-MONU-0019, and US-Israel BSF Grant 2011506.}% <-this % stops a space
\thanks{
Y.C. and S.M. are with the Department of Electrical and Computer
Engg.\ at the University of Florida, Gainesville. A.B.\ is with Inria and the Computer Science Dept. of \'Ecole Normale Sup\'erieure, Paris, France.}%
}

\begin{document}

\maketitle
%\tableofcontents

%%%%%%%%%%%%%%%%%%%%%%%%%%%%%%%%%%%%%%%%%%%%%%%%%%%%%%%%%%%%%%%%%%%%%%%%%%%%%%%%
\begin{abstract} 

It is now well known that flexibility of energy consumption can be harnessed for the purposes of grid-level ancillary services.   In particular,  through distributed control of a collection of loads,   a balancing authority regulation signal can be tracked accurately, while ensuring that the quality of service (QoS) for each load is acceptable \textit{on average}.  In this paper it is argued that a histogram of QoS is approximately Gaussian, and consequently each load will eventually receive poor service.  Statistical techniques are developed to estimate the mean and variance of QoS as a function of the  power spectral density of the regulation signal.  It is also shown that additional local control can eliminate risk:  The histogram of QoS is \textit{truncated} through this local control, so that strict bounds on service quality are guaranteed.   While there is a tradeoff between the grid-level tracking performance (capacity and accuracy) and the bounds imposed on QoS, it is found that the loss of capacity is minor in typical cases.

\spm{QoS may not be something someone would search for in this context.  "demand response" is}

\end{abstract}

%\tableofcontents

%%%%%%%%%%%%%%%%%%%%%%%%%%%%%%%%%%%%%%%%%%%%%%%%%%%%%%%%%%%%%%%%%%%%%%%%%%%%%%%%

%\section*{Todo}

%1. 8-page limit

%2. one paragraph abstract
%spm{to do:  future work,  QoS/2, Regulation signal used to obtain \Fig{fig:cut-off}.}
%3. future work

%Numeric results:

%6. Regulation signal used to obtain \Fig{fig:cut-off}.

%7. QoS intervals in \ref{s:num} have to be replaced by their values/2.

\section{Introduction} 
\label{s:intro}

The power grid requires regulation to ensure that supply matches demand.  Regulation is required by each
 \textit{balancing authority} (BA) on multiple time-scales, corresponding to the time-scales of volatility of both supply and demand for power.  Resources that supply these regulation services are collectively known as  \textit{ancillary services}.  FERC orders 755 and 745 are 
 examples of recent federal policy
 intended to provide incentives for the provision of these services.

A number of papers have explored the potential for extracting ancillary service through the inherent flexibility of loads.
%\spm{Goes back at least to Schweppe}
Examples of loads with sufficient flexibility to provide service to the grid are aluminum manufacturing,
 plug-in electric vehicles,  heating and ventilation (HVAC), and water pumping for irrigation 
\cite{malcho85,matkoccal13,kizmal13,meybarbusyueehr15,haolinkowbarmey14}. 
\spm{I'm not sure about data centers.  I think the HVAC systems at data centers are useful!  I deleted this example.  }
%meybarbusehr13 haokowlinbarmey13
    Even with direct load control, there may be delay and dynamics, so harnessing ancillary services from flexible loads amounts to a control problem: The BA wishes to design some signal to be broadcast to loads, so that deviation in power consumption tracks a reference signal.   It has been argued that a randomized control architecture at each load can simplify this control problem    \cite{meybarbusyueehr15,barbusmey14,YueChenThesis16}.
       \spm{These guys don't do anything clever at the load: matkoccal13, and they don't simplify the control problem!
       \\
       we don't have to impose this architecture as the only way: based on measurements of aggregate power output  }

\begin{figure}[h]
\centering
\Ebox{.65}{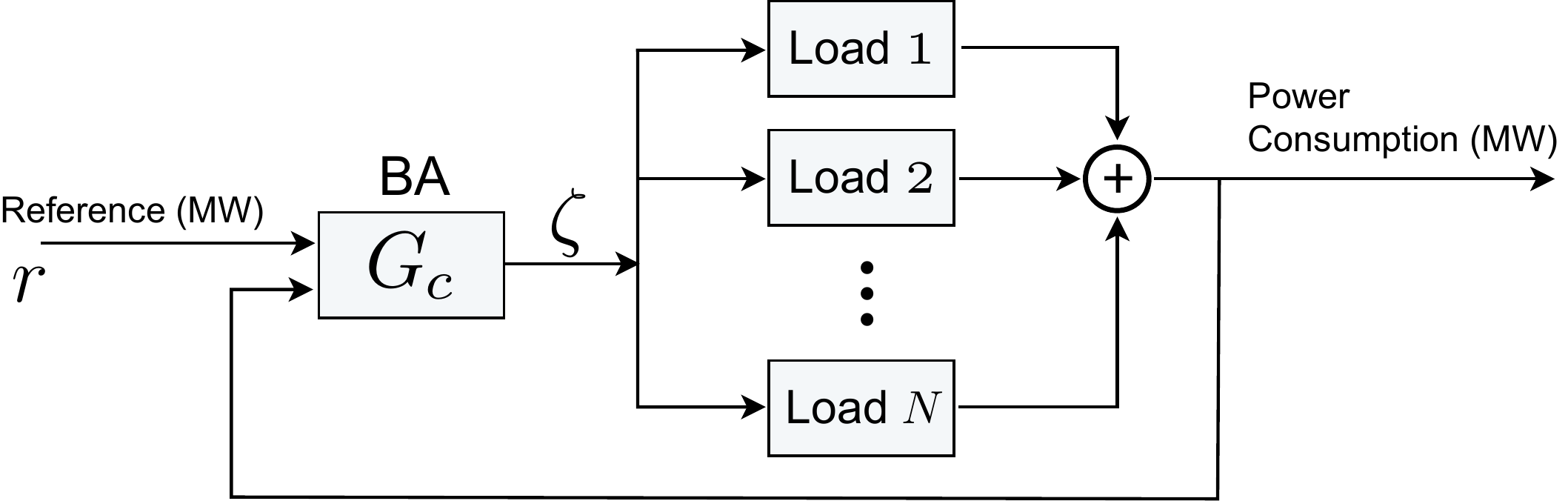} 
\vspace{-.2cm}
\caption{Control architecture for Demand Dispatch} 
\vspace{-.1cm}
% command $\bfzeta$ is computed at a BA, and transmitted to each load. The control decision at a load is based only on its own state and the signal $\bfzeta$.}
\label{fig:arch} 
\end{figure}

\Fig{fig:arch} shows a schematic of the \textit{Demand Dispatch} control architecture adopted in  \cite{meybarbusyueehr15}, 
in which each load operates according to a randomized policy based on its internal state, and a common control signal $\bfzeta$. 
Theoretical results and examples in this prior work demonstrate that local randomized policies can be designed to simplify  control of the aggregate to provide reliable ancillary services.

% {needed?  The analysis was based on a mean field limit (as the number of loads approaches infinity), combined with a linear time-invariant (LTI) system approximation of the aggregate nonlinear model.   In the examples considered in this prior work, the linear approximation was found to be minimum phase, which is why simple linear error feedback could be applied to achieve nearly perfect tracking.  
%}

Absent in prior work is any detailed analysis of risk for an individual load (with the exception of the preliminary work \cite{chebusmey14} on which the present paper is built).  In the setting of  \cite{meybarbusyueehr15} it can be argued that  the quality of service (QoS) for each load is acceptable only \textit{on average}.   Strict bounds on QoS are addressed in \cite{leboudec11} for a deterministic model.  The \textit{service curves} considered there are of similar flavor to the QoS metrics used in the present work.  

%The resulting input-output model is \textit{positive real} under general conditions, and was found to be minimum phase in most examples.

%This paper examines in depth the issue of individual risk. 
The main contributions of this paper are summarized as follows:  QoS metrics are proposed in the first part of the paper,  along with techniques to approximate their mean and variance.   Estimates of second order statistics are a function of the  power spectral density (PSD) of the grid regulation signal.  These theoretical results are developed in \Section{s:MFMload}.

A simple approach is proposed to restrict QoS to pre-specified bounds:  A load will opt-out of service to the grid temporarily, whenever its QoS is about to exit pre-specified bounds.   This essentially eliminates risk since the histogram of QoS is restricted to these bounds.

 \begin{figure}[h]
 \centering
\Ebox{.75}{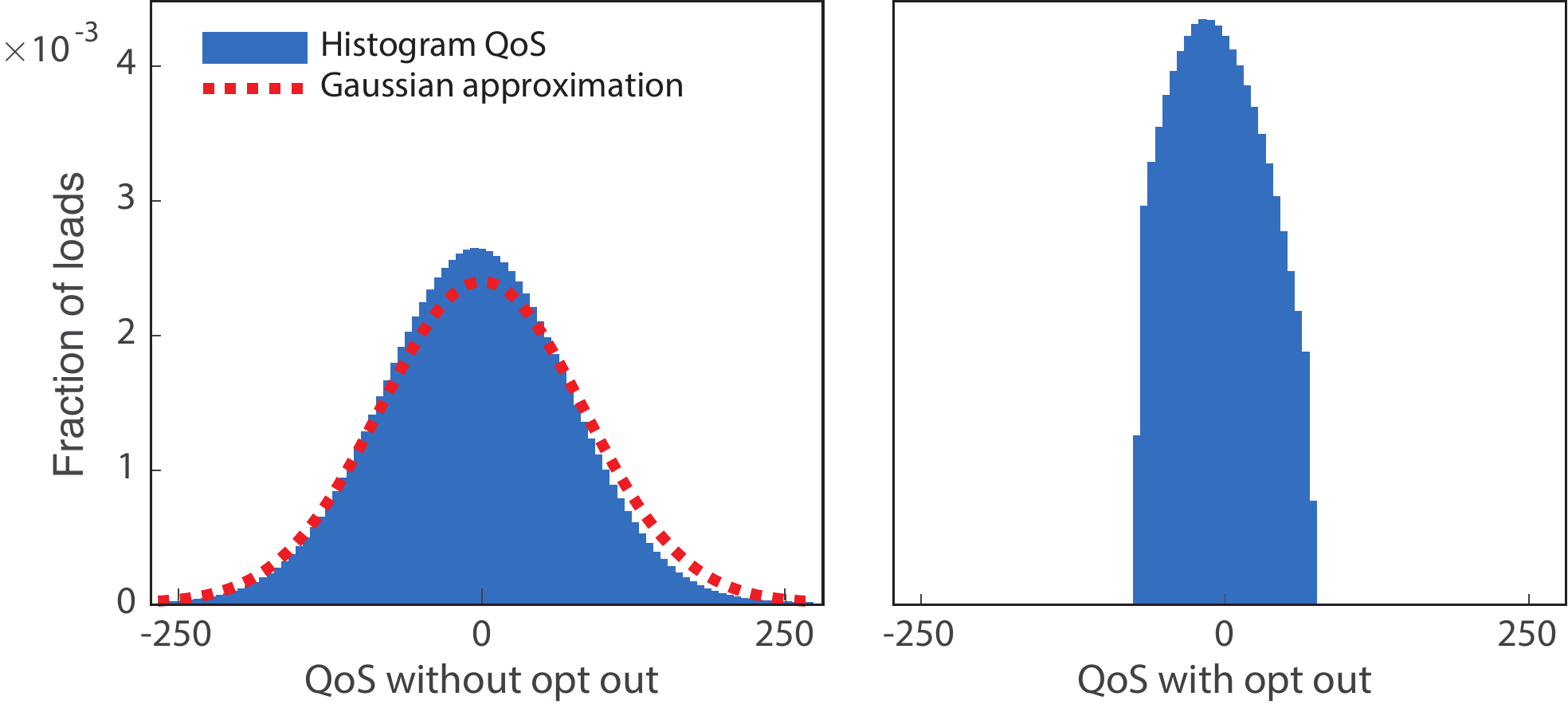}
\vspace{-.2cm}
\caption{Discounted QoS \eqref{e:Health} with and without local opt-out control.}
\label{fig:BothHist} 
\end{figure} 
 
Numerical results surveyed in \Section{s:num} confirm that the histogram of QoS is truncated through this local control, so that strict bounds on service quality are guaranteed. As long as the QoS bounds are not overly restrictive, the  impact on the grid-level performance is \textit{insignificant}.

\Fig{fig:BothHist} shows histograms of QoS based on simulation experiments described in \Section{s:num}.  The plot on the left hand side shows that a Gaussian approximation is a good fit with empirical results when there is no local opt-out control.  The figure on the right shows how the histogram is truncated when opt-out control is in place.

The methodology used in this work employs a linear representation of the controlled Markov chain that represents the state of an individual load \cite{chebusmey16e}.   Similar representations are used in \cite{lipkrirub84,chebusmey15b} to obtain nonlinear filters  for state estimation.

We begin with a brief survey of a portion of results from \cite{meybarbusyueehr15},  and a precise definition of QoS for a load. 

\section{Randomized control and mean-field models} 
\label{s:mfm}

\subsection{Randomized control}

The system architecture considered in this paper is
illustrated in \Fig{fig:arch}, based on the following components: 
\begin{romannum}
\item 
There are $N$ homogeneous loads that receive a common scalar command signal from the balancing authority (BA), denoted $\bfzeta=\{\zeta_t\}$ in the figure.

\item
Each load evolves as a controlled Markov chain on the finite state space $\state=\{x^1,\dots,x^d\}$.
Its transition probability is determined by its own state, and the BA signal $\bfzeta$. The common dynamics are defined by a controlled transition matrix $\{P_\zeta : \zeta\in\Re\}$. For the $i$th load, there is a state process $\bfmX^i$ whose transition probability is given by,
\begin{equation}
\Prob\{X^i_{t+1} = x' \mid X^i_r, \zeta_r : r\le t \} = P_{\zeta}(x,x') ,  
\label{e:Pzeta}
\end{equation}
for each $x'\in\state$,  $X^i_t = x\in\state$ and $\zeta_t=\zeta\in\Re$.

\item
The BA has measurements of the other two scalar signals shown in the figure: The aggregate power consumption $\bfmy$ and desired deviation power consumption $\bfmr$.

\end{romannum}
An approach to construction of $\{P_\zeta:\zeta\in\Re\}$ was proposed in  \cite{meybarbusyueehr15} based on information-theoretic arguments. The \emph{nominal behavior} is defined as the dynamics with $\bfzeta\equiv 0$.  

In the present paper we do not require a specific construction of $P_\zeta$, but the following assumptions are imposed in our main results. 
\begin{romannum}
\item[\textbf{A1:}]
 The transition matrix $P_\zeta$ is twice continuously differentiable ($C^2$) in a neighborhood of $\zeta=0$,   and   the second derivative is Lipschitz continuous. In addition, the nominal transition matrix $P_0$ is irreducible and aperiodic.

The first and second order derivatives of the transition matrix   at $\zeta = 0$ are denoted,
\begin{equation}
\clE=\frac{d}{d\zeta} P_\zeta \Big|_{\zeta=0},
\qquad
\clE^{(2)}=\frac{d^2}{d\zeta^2} P_\zeta \Big|_{\zeta=0}.
\label{e:Pder}
\end{equation}

\item[\textbf{A2:}]
$\zeta_t = \epsy \zeta_t^1$, where $0 \le \epsy < 1$ and $\bfzeta^1=\{\zeta_t^1 : t \in \intgr\}$ is a   real-valued stationary stochastic process with zero mean. The following additional assumptions are imposed:
\begin{romannum}
\item It is bounded,  $|\zeta_t^1|\le 1$ for all $t$ with probability one.  Hence      $\sigma_{\zeta}^2=\Expect[(\zeta_t)^2]\le \epsy^2$.
\item  Its auto-covariance satisfies, for each $t$,
\[
|\Sigma_{\zeta}(t)| \le \epsy^2 b \rho^{|t|} \text{, with } b<\infty \text{, and } |\rho|<1.
\]

\end{romannum}  
\end{romannum}

It is assumed that the power consumption at time $t$ from load $i$ is equal to some function of the state, denoted $\util(X^i_t)$.   The normalized power consumption is denoted,
\begin{equation}
y_t^N = \frac{1}{N}\sum_{i=1}^N \util(X^i_t).
\label{e:outputN}
\end{equation}
%The superscript is dropped unless dependency on $N$ must be emphasized.

Under Assumption \textbf{A1},  $P_0$ has a unique pmf (probability mass function) $\pi_0$.  The value $\bary^0\eqdef \sum_x \pi_0(x)\util(x)$ is interpreted as the average nominal power usage.  On combining the ergodic theorem for Markov chains with the Law of Large Numbers for i.i.d.\ sequences we can conclude that $y^N_t\approx \bary^0$ when both $N$ and $t$ are large,  and $\bfzeta\equiv 0$.  
 
It is assumed that the  signal $\bfmr$ is also normalized so that tracking amounts to choosing the signal $\bfzeta$ so that $ \tily_t^N \approx r_t$ for all $t$, where $\tily_t^N = y_t^N - \bary^0$ is the deviation from nominal behavior. For example, we might use error feedback of the form, 
\begin{equation}
\zeta_t = G_c e_t,\qquad e_t = r_t-\tily_t^N   \, ,  
\label{e:zetaGc}
\end{equation} 
where $G_c$ is the control transfer function
\cite{coupertemdeb12,macalhis10,meybarbusyueehr15}.    Adopting terminology from control engineering,  $\bfmr$ will be called the \textit{reference signal}.
\spm{Yue:  Note!  You will use "slash-defn{reference signal} in your thesis.  
\\
 Convention after discussion on May 27:  reference refers to a signal we want a collection of loads to track.   }

\subsection{Mean-field model}

The mean-field model is based on  the empirical pmfs:
\begin{equation}
\mu^N_t(x)\eqdef \frac{1}{N}\sum_{i=1}^N  \ind\{ X^i_t =x \} ,\quad x\in\state.
\label{e:empDist}
\end{equation}
Each entry of the vector  $\mu_t^N$   represents the fraction of loads in a particular state.  Under very general conditions on the input sequence $\bfzeta$,  it can be shown that the empirical distributions converge to a solution to the nonlinear state space model equations, 
\begin{equation}
\mu_{t+1} = \mu_t P_{\zeta_t},
\label{e:MFM}
\end{equation} 
where $\mu_t$ is considered as a row vector. 
The output is denoted $y_t=\sum_x \mu_t(x)\util(x)$, which is the limit of the $\{y_t^N \}$ (defined in \eqref{e:outputN}) as $N\to\infty$.  See
 \cite{meybarbusyueehr15} for details.

The unique equilibrium with $\bfzeta\equiv 0$ is $\mu_t\equiv \pi_0$ and $y_t\equiv \bary^0 $.  The analysis in \cite{meybarbusyueehr15} can be extended to obtain a linearization about this equilibrium.  This is described by the linear state space model,
\begin{equation}
\begin{aligned}
 \Phi_{t+1} &= A \Phi_t + B \zeta_t
 \\
\gamma_t &= C \Phi_t
\end{aligned}
\label{e:LSSmfg}
\end{equation}
where
$A=P^\transpose_0$, $C$ is a row vector of dimension $d=|\state|$ with $C_i= \util(x^i) $ for each $i$, and $B$ is a $d$-dimensional column vector with entries $B_j = \sum_x\pi_0(x) \clE(x,x^j) $, where
the matrix $\clE$ is defined in \eqref{e:Pder}.
In the state equations \eqref{e:LSSmfg},  the state  $\Phi_t$ is $d$-dimensional,  and $\Phi_t(i)$ is intended to approximate $\mu_t(x^i) -\pi_0(x^i)$ for $1\le i\le d$.   The output $\gamma_t$ is an approximation of $\tily_t = y_t - \bary^0$.

\subsection{QoS for an individual and the population}
\label{s:QoS}

The QoS metrics considered in this paper are defined by a scalar-valued function $ \health:\state \to \Re$, and a stable transfer function $H_\Health$. The QoS of the $i$th load at time $t$ is then defined by $\Health_t^i = H_\Health\, L^i_t$, where $L_t^i = \health(X_t^i)$. For example, the function $\health$ may represent temperature, cycling, or power consumption as a function of $x\in\state$.

Two classes of transfer functions will be considered in numerical experiments:
\begin{romannum}
	\item Summation   over a finite time horizon $T_f$:
	\begin{equation}
	\Health_t^i =  \sum_{k=0}^{T_f}  \health(X_{t-k}^i).
	\label{e:HealthFinite0}
	\end{equation}
	
	\item Discounted sum:
	\begin{equation}
	\Health_t^i  = \sum_{k=0}^\infty \beta^k \health(X_{t-k}^i)\,,
	\label{e:Health}
	\end{equation}
	where  the discount factor satisfies $\beta\in[0,1)$.  
\end{romannum}
In particular,   setting $T_f = 0$ or $\beta =0$ gives $\Health_t^i = \health(X_t^i)$.

\spm{avoid text without content!   that only depends on the load state at time $t$.}

Unless elsewhere specified, the function $\health$  is specialized to reflect the power consumption of a load,
\begin{equation}
\health(X_t^i) = \util(X_t^i),
\label{e:health_power}
\end{equation}
and its normalized form is also considered,
\begin{equation}
\health(X_t^i) = \util(X_t^i) - \bary^0
\label{e:health_Npower}
\end{equation}
where $\bary^0$ is defined after \eqref{e:outputN}.

The average  QoS at time $t$ is denoted, 
\[
\bar{\Health}_t = \frac{1}{N}\sum_{i=1}^{N}\Health_t^i\,,
\]  
and the filtered signal is denoted $R_t= H_\Health  r_t$.
The following result follows from the definitions:
\begin{proposition}
	\label{t:H_approx}
	Suppose there is perfect tracking: $\tily_t^N=r_t$ for all $t$. Then, under the definition of $\health(X_t^i)$ in \eqref{e:health_Npower},
	\begin{equation}
	\bar{\Health}_t = R_t
	\label{e:Health=R}.
	\end{equation}
	\qed
\end{proposition}

%%\begin{proof}
%\textit{Proof:} Based on the definition \eqref{e:outputN}, the averaged value $\bar{L}_t= \frac{1}{N} \sum_{i=1}^N \health(X_t^i)$ is equal to the normalized power deviation $\tily_t^N$.
%If $\tily_t^N = r_t$ as assumed in the proposition, then $\bar{L}_t= r_t$. The desired conclusion \eqref{e:Health=R} is obtained since $\bfmR= H_\Health \, \bfmr$ and $\bar{ \bfmath{\Health}} = H_\Health \, \bar{\bfmL}$. 
%\qed

In practice we can only expect the approximation $\tily_t^N\approx r_t$,  which will imply the corresponding approximation $	\bar{\Health}_t \approx R_t$.  This leads to a useful heuristic notion of capacity of the aggregate of loads as a function of time:

\paragraph*{Battery analogy}
With $\health$ defined in \eqref{e:health_Npower}, the signal $\bar{\Health}_t$ is similar to the SOC (state of charge) of a  battery. In particular, a large value of $\bar{\Health}_t$ suggests loads have a large capacity for ``discharge". Since  $R_t$ approximates $\bar{\Health}_t$ by design, the former can be used as an indicator of  SOC of loads.

\subsection{Example: Intelligent pools}
\label{s:pool}

The paper \cite{meybarbusyueehr15} on which the present work is based considered an application of this system architecture in which the loads are a collection of pools.  The motivation for considering pools is the inherent flexibility of pool cleaning, and because the total load in a region can be very large.  The maximum load is approximately 1GW in the state of Florida.   

In this prior work,   the state space was the finite set,
\begin{equation}
\state=\{ (\kappa,j) : \kappa \in \{ \oplus,\ominus\} ,\ j\in \{1, 2, \dots , \clI \} \}. 
\label{e:poolstate}
\end{equation}
The load samples the grid signal periodically  (the sampling increments are assumed to be deterministic, or i.i.d.\ and distributed according to a geometric distribution).  At the time of the $k$th sample, if $X_k = (\ominus,j)$ then the load   has remained off for the past $j$ sampling times,  and was turned off at sampling time $k-j$.  The interpretation of $X_k = (\oplus,j)$ is symmetrical, with ``$\oplus$'' indicating that the load is currently consuming power.

A technique to construct the state transition matrix $P_\zeta$  was introduced in \cite{meybarbusyueehr15} using an optimal-control approach, and improved in \cite{chebusmey15b,busmey16v} to reduce the dimension of the state space,  and extend application beyond this class of loads.
It has been demonstrated through extensive simulation that tracking at the grid level is nearly perfect, provided that the reference signal is scaled and filtered \cite{chebusmey14,meybarbusyueehr15,chebusmey15b}. 
What was left out in prior work is any detailed consideration of the quality of service to individual loads.  The pool pump example is ideal for illustrating the possibility of risk for an individual load, and illustrating how this risk can be reduced or even eliminated.

 %\spm{We claim 100,000 pools in simulation, but actually, we simulated with 99,996 pools to make it the multiple of m = 6. }
 
\begin{figure}[h]
  \centering
\Ebox{.65}{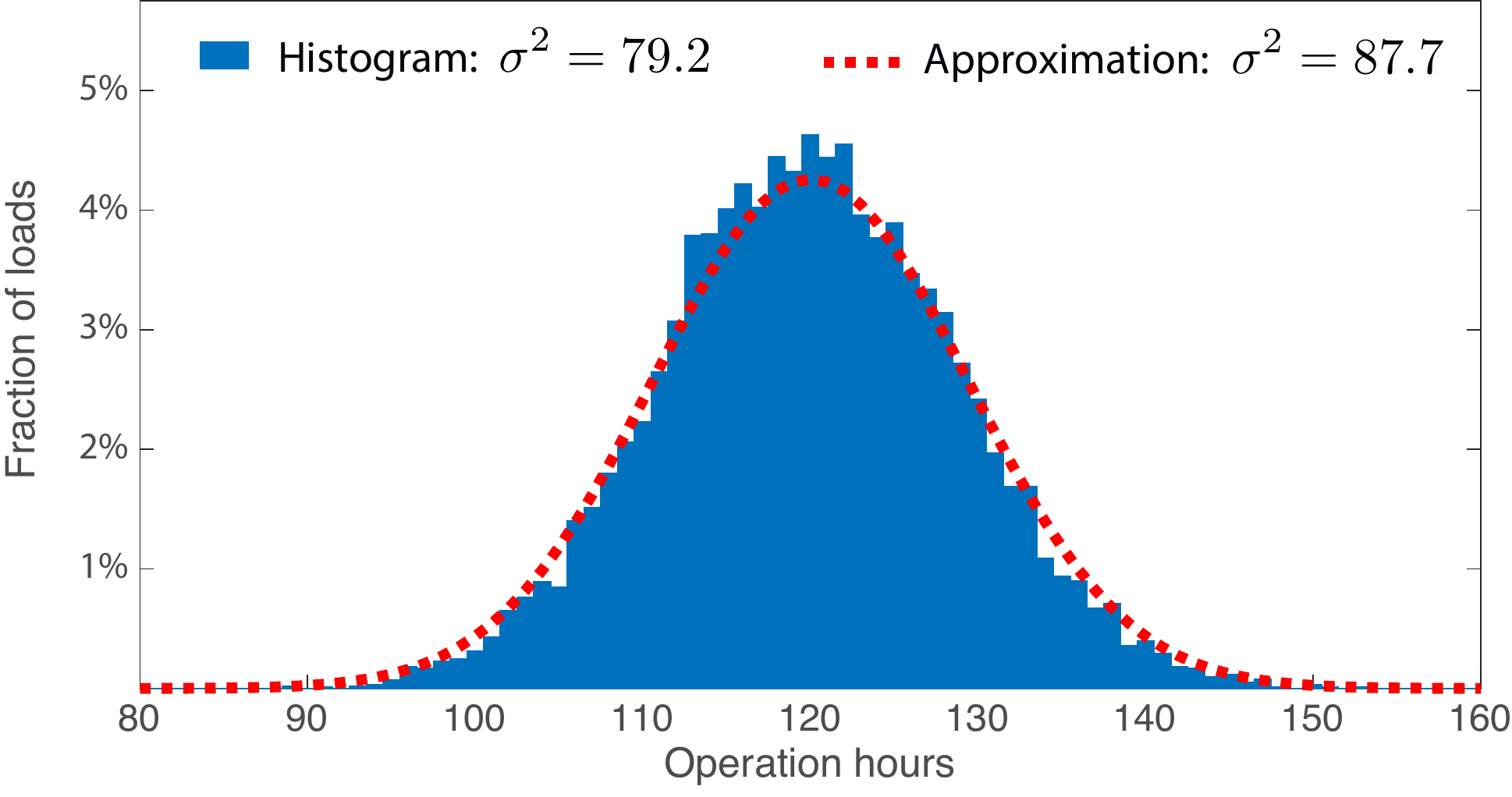}
\vspace{-.2cm}
\caption{Histogram of the moving-window QoS metric \eqref{e:HealthFinite0}.}
\vspace{-.1cm}
\label{fig:10days_hist}
\end{figure}

In a numerical experiment conducted with $10^4$ pools, each pool pump was expected to clean the pool 12 hour/day and assumed to consume 1 kW when operating.  Hence the function $\util$ that defines \eqref{e:health_power} and \eqref{e:health_Npower} is the indicator function $\util(x) = \sum_j \ind\{x = (\oplus, j) \}$.
\Fig{fig:10days_hist} shows the histogram of the QoS metric $\Health_t^i$ based on \eqref{e:HealthFinite0}  and  \eqref{e:health_power},
with $T_f$ corresponding to 10 days.

The histogram appears Gaussian, with a mean of approximately  120 hours ---  this is consistent with the time horizon used in this experiment, given the nominal 12 hour/day cleaning period.  It is evident that a  fraction of pools are over-cleaned or under cleaned by 24 hours or more. 
 
An analysis of QoS is presented in the next section based on a model of an individual load in the mean field limit.

\section{QoS analysis and opt-out control}
\label{s:MFMload}

In the mean-field limit, the aggregate dynamics are deterministic, following the discrete-time nonlinear control model \eqref{e:MFM}. The behavior of each load remains probabilistic. 

\subsection{Mean field model for an individual load}

The mean field model for \textit{one load} is defined by replacing $(\mu^N_t,\zeta^N_t)$ with its mean-field limit $(\mu_t,\zeta_t)$. The justification is that we have a very large number of loads, but our interest is in the statistics of an individual. 
\spm{Note that we cannot state that $\bfzeta$ is deterministic since we discuss its PSD}

%Under the conditions under which the mean field model is obtained as a limit,  the signal $\bfzeta$  is a function of the reference signal $\bfmr$ that can be expressed as causal feedback,
%\[
%\zeta_t = \phi_t(\mu_0,\dots,\mu_t, r_0,\dots, r_t),\qquad t\ge 0,
%\]
%where in the mean field model, the sequence of pmf evolve according to \eqref{e:MFM}.
%Hence $\zeta_t $ is a nonlinear function of $r_0,\dots, r_t$, and the initial condition $\mu_0$ 
%(which is assumed to be deterministic).   

%Letting $Z$ denote the $z$-transform of $\{\zeta_t\}$,
%and
%$R$ denote the $z$-transform of $\{r_t\}$, we use the approximation,
%\begin{equation}
%Z= \frac{G_c}{1+G_cG_p} R
%\label{e:zetainfinity}
%\end{equation}

The super-script $i$ is dropped in our analysis of a single load. Hence $\bfmX$ denotes the controlled Markov chain whose transition probabilities are defined consistently with \eqref{e:Pzeta}:
\begin{equation}
\Prob\{X_{t+1} = x' \mid X_r, \zeta_r : r\le t \} = P_{\zeta_t}(x,x') ,\quad x=X_t
\label{e:Pzetab}
\end{equation}
The construction of the mean field model \eqref{e:MFM} is based on lifting the state space from the $d$-element set
$\state = \{x^1,\cdots,x^d\}$, to the $d$-dimensional simplex $\Spx$. For the $i^{th}$ load at time $t$, the element $\piload_{t} \in \Spx$ is the degenerate distribution whose mass is concentrated at $x$ if $X_t= x$;
that is, $\piload_{t} =\delta_x$.

With this state description, the load evolves according to a random linear system, similar to   \eqref{e:MFM}:
\begin{equation}
	\piload_{t+1}=\piload_{t}G_{t+1} 
\label{e:piG}
\end{equation}
in which $\piload_{t}$ is interpreted as a $d$-dimensional row vector.   The  $d\times d$ matrix 
$G_{t}$ has entries $0$ or $1$ only, with $\sum_{x'\in\state} G_{t}(x,x')=1$ for all $x\in\state$. It is conditionally independent of $\{\piload_0,\cdots,\piload_{t}\}$, given $\zeta_t$, with
\begin{equation}
\label{e:EG=P} 
 	\Expect[G_{t+1}|\piload_0, \cdots, \piload_{t}, \zeta_t]=P_{\zeta_t}.
\end{equation}

The random linear system \eqref{e:piG} can be described as a linear system driven by ``white noise'':
\begin{equation}
	\piload_{t+1}=\piload_{t}P_{\zeta_t}+\Delta_{t+1}
	\label{e:Sys_delta}
\end{equation}
where, $\{\Delta_{t+1}=\piload_{t}(G_{t+1} -P_{\zeta_t}): t\geq 0 \}$ is a martingale difference sequence.

A Taylor-series approximation of $P_{\zeta_t}$  leads to a useful approximation of \eqref{e:Sys_delta}.  
Recall that the first and second derivatives $\clE$ and $\clE^{(2)}$ were introduced in \eqref{e:Pder}.
The proof follows from the definitions.

%Applying the first and second derivatives of $P_{\zeta_t}$ in \eqref{e:Pder},  the approximation of \eqref{e:Sys_delta} follows in \Proposition{t:LTI approx}.

\begin{proposition}
\label{t:LTI approx}
The nonlinear system \eqref{e:Sys_delta} admits the LTI approximation,
\begin{equation}
	\piload_{t+1}=\piload_{t}P_0+D_{t+1}+O(\epsy^3)
	\label{e:LTI approx}
\end{equation}
in which, 
\begin{equation}
D_{t+1} \eqdef
B_{t}^\transpose \zeta_t +  V_t^\transpose \zeta_t^2 +\Delta_{t+1},
\label{e:D}
\end{equation}
with
$B_{t}^\transpose =\piload_{t} \clE$ and $V_t^\transpose = \frac{1}{2}\piload_t \clE^{(2)}$.
\end{proposition}

\subsection{Steady-state QoS statistics}
\label{s:QoS_stat}

The main goal of this section is to estimate the second order statistics of $\{\Health_t\}$ for arbitrary function 
$\health$ and stable filter  $H_\Health$.   These approximations are obtained for a stationary realization.

 \begin{theorem}
\label{t:stationary}
 Under assumptions \textbf{A1} and \textbf{A2},
there exists a realization $\{\zeta_t, X_t, D_t, \Health_t: -\infty < t < \infty \}$ that is jointly stationary.
 \qed
\end{theorem}

The proof of the existence of the stationary realization $\{\zeta_t, X_t, D_t : -\infty < t < \infty \}$ follows from Proposition~2.3 of  \cite{chebusmey16e}.    Stationarity of the joint process $\{\zeta_t, X_t, D_t, \Health_t\}$ follows from the assumption that $ \{\Health_t\}$ is obtained from $\{\ell(X_t)\}$ through a stable filter.  
Proposition~2.5 of \cite{chebusmey16e}  implies the following  approximation for the mean QoS.    

%contains a second order Taylor series approximation of the steady state mean $\Expect[\piload_t]$. Ignoring the second-order term gives

\begin{proposition}
The mean QoS admits the approximation,
\[
\Expect[\Health_t] = H_\Health(1) \Expect[L_t], \quad \Expect[L_t] = \sum_x \health(x)\pi_0(x) + O(\epsy^2)
\]
where $H_\Health(1)$ is the DC gain of the transfer function $H_\Health$.
\qed
\end{proposition}

The PSD for the stationary realization of the stochastic process $\bfmD$ can be approximated by a second order Taylor series expansion:  Applying Theorem 2.4 of \cite{chebusmey16e}, a function $\psdder_D \colon [-\pi,\pi] \to \Co^{d\times d}$ can be constructed such that 
\begin{equation}
\psd_D(\theta) = \psd_D^\bullet(\theta) + \epsy^2 \psdder_D(\theta) + o(\epsy^2),\quad \theta\in [-\pi,\pi] \,,
\label{e:psd_taylor}
\end{equation}
in which $\psd_D^\bullet$ is the PSD obtained with $\bfzeta\equiv 0$.  This is independent of $\theta$ since $\bfmD$ is a white noise sequence in this case.   An approximation for the PSD of $\{\Health_t\}$ follows from \eqref{e:psd_taylor}
 and the following:
\begin{romannum}
\item 
The model \eqref{e:piG} is approximated by the linear state space model, 
\begin{equation}
\widehat{\piload}_{t+1} = \widehat{\piload}_t P_0 +  D_{t+1}\, .
\label{e:Sys_delta_lin}
\end{equation}

\item
Formulae for $ \psd_D^\bullet$ and $ \psdder_D$ are given in \cite{chebusmey16e}.
The computation of $\psd_D^{(2)}$ in \eqref{e:psd_taylor} requires the autocovariance of $\bfzeta$.
This must be estimated from data.

% which is assumed in \textbf{A2}.  An example of constructing the $\bfzeta$ model is provided in\Section{s:zeta_model}.

\item
The approximation of the PSD $\psd_{\Health}$  for $\{\Health_t\}$ is obtained using \eqref{e:psd_taylor}
 and the linear model shown in \Fig{f:health}.
\end{romannum}

The variance is then obtained based on an inverse formula, or an approximation:

\begin{figure}
\centering
\Ebox{.65}{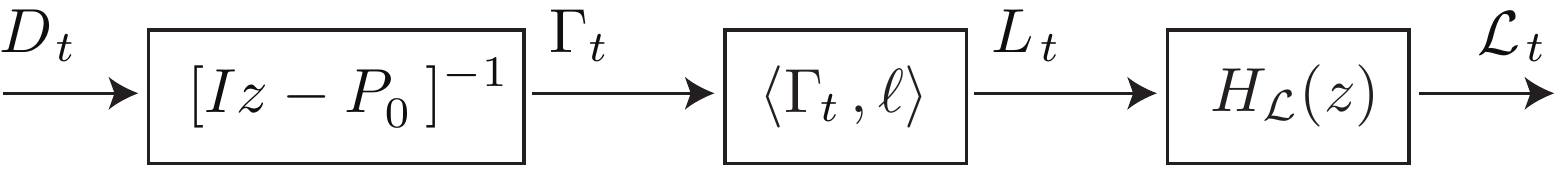}
\vspace{-.2cm}
\caption{Linear model of QoS evolution. } 
\label{f:health}
\end{figure}

\begin{theorem}
\label{t:var_taylor}
The variance of QoS admits the following representation and approximations:
\begin{romannum}
\item 
The variance of QoS is the average of its PSD:
 \begin{equation}
{\clV}_{\Health} = \frac{1}{2\pi} \int_0^{2\pi} \psd_{\Health}(\theta) \, d\theta.
\label{e:VarQoS}
\end{equation}  

In the special case of the moving time-horizon \eqref{e:HealthFinite0}, it admits the approximation
\begin{equation}
\frac{\clV_\Health}{T_f} \approx \psd_{L}(0) \quad \text{if $T_f \approx \infty$,}
\label{e:VarQoSapprox}
\end{equation}
where $\psd_{L}$ is the PSD of $\bfmL$.

\spm{Yue, careful with periods after equations when there is text following}
\item  A Taylor series of the right hand side of  
\eqref{e:VarQoS} implies the approximation,
\begin{equation}
{\clV}_{\Health} = {\clV}_{\Health}^\bullet+ \epsy^2 {\clV}_{\Health}^{(2)} + o(\epsy^2),
\label{e:var_taylor}
\end{equation}
where the terms ${\clV}_{\Health}^\bullet$ and ${\clV}_{\Health}^{(2)}$ are obtained using  the QoS variance formula in \eqref{e:VarQoS},  and the approximation of the PSD $\psd_\Health$ is obtained using \eqref{e:psd_taylor}.
%, along with the PSD $\psd_\Health$ that based on the approximation for $\psd_D$ given in \eqref{e:psd_taylor} and the linear mapping: $\bfmD \to \bf{\Health}$ described in \Fig{f:health}.
\qed
\end{romannum}
\end{theorem}

Recall that the QoS histogram  shown in \Fig{fig:10days_hist} is based on the  moving time-horizon \eqref{e:HealthFinite0},   with $T_f$ corresponding to 10 days.  The approximation of the QoS variance was obtained using \eqref{e:VarQoSapprox} and a Taylor series approximation of $ \psd_{L}(0)$.

\subsection{Opt-out control}
\label{s:optout}

An extra layer of control is required to truncate the  two tails of the QoS histograms observed in experiments.

The local control considered in this paper is a simple ``opt-out'' mechanism, based on pre-defined upper and lower limits  $b_+$ and $b_-$.   A load ignores a command from grid operator at time $t$ if it may result in $\Health_{t+1}\not\in [b_-,b_+]$, and  takes an alternative action so that $\Health_{t+1} \in [b_-,b_+]$ with probability one.  
\spm{added since in some choices of $\ell$ there may be some uncertainty}
This ensures that the QoS metric of each load remains within the predefined interval for all time.

In many cases, the poor QoS revealed by the two tails of the histogram represents only a small portion of loads. 
Therefore, the impact from local opt-out control is insignificant at the grid level if the QoS interval $[b_-,b_+]$ is carefully chosen.

%  Not sure about this:
%Based on all computations above , we obtain the variance of load health value $\sigma_{\Health}^2=183$ corresponding to the reference signal in \Fig{fig:0LC_output}. The Gaussian distribution with mean zeros and variance $\sigma_{\Health}^2 $ was plotted as the dashed line in \Fig{fig:0LC_hist}. The approximation is close to the simulation result.
%%\subsection{Variance estimates}

%\begin{figure} 
%\Ebox{.85}{3_0LC_refx1_Output.pdf}
%\vspace{-.2cm}
%\caption{Tracking performance without local control.}
%\label{fig:0LC_output}
%\vspace{-.2cm}
%\end{figure} 

\section{Numerics}
\label{s:num}

Numerical experiments were conducted on the pool pump model to illustrate the main technical conclusions. Simulation results are summarized below: 
\begin{romannum}

\item
Applications of \Theorem{t:var_taylor} show that the approximations of QoS closely match observed QoS. 

\item  Opt-out control ensures that QoS lies within strict bounds, and  tracking remains nearly perfect in most cases. In some extreme cases, capacity is reduced with the introduction of opt-out control.  However, it is found through simulations that it is far less than might be predicted by the approximation in (i).

\item  The opt-out control can be applied to multiple QoS metrics.
The capacity is further reduced with the introduction of an additional QoS metric, but the reduction is found to be minor numerical experiments.

\item   Approaches are proposed and tested in \Section{s:num_SOC} to condition the reference signal to reduce potential capacity reduction caused by opt-out local control. 
\end{romannum}

%(v) 

\subsection{Simulation setup}
\label{s:num_setup}

The simulation used $N=10^4$ homogeneous Markov models:  
Each pool pump is operated under a 12 hours/day cleaning cycle, and consumes $1$~kW during operation.

Two QoS metrics are considered, differentiated by the function $\ell$ appearing in the definition $L_t ^i= \health(X_t^i)$:
In the first QoS function, the normalized power consumption \eqref{e:health_Npower} is considered so that
if $\Health_t^i > 0$ ($\Health_t^i < 0$) then the pool has been over-cleaned (under-cleaned).

The second QoS function is introduced to capture the on/off cycling of loads:  $ \health^c(X_t^i,X_{t+1}^i) 
 =  $
\begin{equation}
\begin{aligned}
 &\sum_j   \Bigl\{ | \ind\{X_{t+1}^i= (\oplus, j) \} - \ind\{X_t^i = (\ominus, j) \}| 
\\
& \qquad  + | \ind\{X_{t+1}^i= (\ominus, j) \} - \ind\{X_t^i = (\oplus, j) \} | \Bigr\}.
\end{aligned}
\label{e:health_swt}
\end{equation}
  These two QoS metrics can be applied to many other loads, such as air conditioners, refrigerators, and water heaters~\cite{busmey16v, matkadnusmey16}. 
Consideration of the QoS metric \eqref{e:health_swt} is postponed to \Section{t:2QoS}.

The discounted sum \eqref{e:Health} was used to define $\Health_t^i$ in these experiments.  
The following interpretation is used to obtain insight on the choice of the discount factor  in this QoS metric. 

Let $\xi$ denote a random variable that is independent of $\bfmX^i$. Its distribution is taken to be geometric on $\nat$, with parameter $\beta$. Its mean is thus
\begin{equation}
\Expect[\xi] = \sum_{k=0}^{\infty} \Prob(\xi \ge k) = \sum_{k=0}^{\infty} \beta^k = \frac{1}{1-\beta}.
\label{e:geo}
\end{equation}
By independence we also have,
\begin{align*}
\Expect \Bigl[\sum_{k=0}^\xi \health(X_{t-k}^i) \Bigr] 
& = \sum_{k=0}^\infty \Expect[ \health(X_{t-k}^i) \ind (\xi \ge k)] 
\\
&= \sum_{k=0}^\infty \Expect[ \health(X_{t-k}^i)] \Prob (\xi \ge k)
\\
&= \sum_{k=0}^\infty \Expect[  \health(X_{t-k}^i)]  \beta^k  
	= \Expect [\Health_t^i] 
\end{align*}

Hence the discounted sum \eqref{e:Health} is similar to the moving window QoS metric \eqref{e:HealthFinite0}. In the following experiments we took $\Expect[\xi] = 10$~days, which corresponds to 2880 time samples (given the $5$ minute sampling period). Solving \eqref{e:geo} gives   $\beta =1-1/2880$.   
%  \approx 0.99965$.  Don't see any value for this approximation.
\spm{Yue, never put words in math mode.  $days$ is ugly!}

\subsection{Model for the stationary input}
\label{s:zeta_model}

\begin{figure}
\centering
	\Ebox{.55}{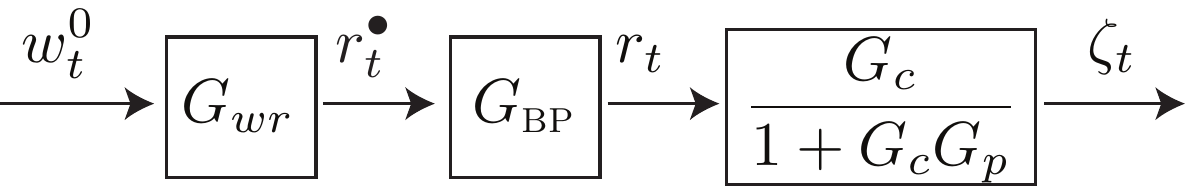}
	\vspace{-.2cm}
	\caption{The input $\bfzeta$ modeled as a stationary stochastic process}
	\vspace{-.4cm}
	\label{fig:zetamodel}
\end{figure}

A linear model for the stationary input $\bfzeta$ was constructed based on the block diagram shown in \Fig{fig:zetamodel}. As in the prior work \cite{meybarbusyueehr15,chebusmey14}, it is assumed that the signal $\bfmr$ is obtained by filtering the regulation signal $\bfmr^\bullet$;  the latter is modeled as filtered white noise with transfer function $G_{wr}$.  
\spm{regulation OK here!!}

The BPA (Bonneville Power Authority~\cite{BPA}) balancing reserves, deployed in January 2015, were taken for the regulation signal $\bfmr^\bullet$. The sampling time is 5 minutes.

\begin{figure}[h]
\centering
\Ebox{.75}{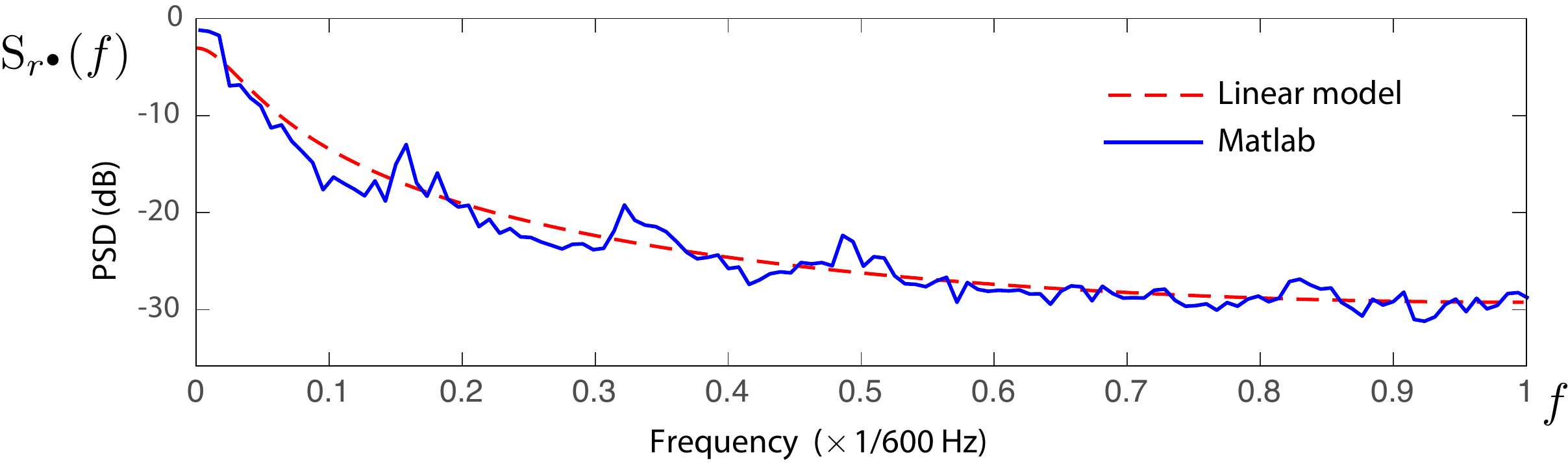}
\vspace{-.2cm}
\caption{BPA signal:  Power spectral density  and its approximation.} 
\vspace{-.1cm}
\label{f:BPA}
\end{figure} 

It is assumed that $\bfmr^\bullet$ evolves as the ARMA (autoregressive moving average) model,
\begin{equation}
	r_t^\bullet +a_1r_{t-1}^\bullet+a_2r_{t-2}^\bullet=w_t+b_1w_{t-1}
 \label{e:ARMA}
\end{equation}
in which $\bfmw$ is white noise with variance $\sigma_w^2$. The extended least squares (ELS) algorithm was used to estimate the coefficients $a_1$, $a_2$, $b_1$, and the variance $\sigma_w^2$ based on the BPA balancing reserves. The algorithm terminated at $[a_1, a_2, b_1]^T= %[-0.9009, 0.0365, 0.0859]^T 
[-1.16, 0.2301, -0.2489]^\transpose$, and $\sigma_w^2=4.36\times 10^{-3}$. In the $z$-domain, its transfer function is expressed
\begin{equation}
%	G_{wr}(z)=\frac{1+0.08594z^{-1}}{1-0.9009z^{-1}+0.03653z^{-2}}.
	G_{wr}(z)=\frac{1-0.2489z^{-1}}{1-1.6z^{-1}+0.2301z^{-2}}.
 \label{e:G_{BPA}(z)}
\end{equation}
%The dashed line in \Fig{f:BPA} is the estimate of the spectrum given by $ |R^\bullet(e^{jw})|^2 = \sigma_w^2 |G_{wr}(e^{jw})|^2$, and the solid line is the PSD estimated using Matlab's psd command.
The dashed and solid lines in \Fig{f:BPA} represent the estimates of the spectrum given by $ |\psd_{r^\bullet}(e^{j\theta})|^2 = \sigma_w^2 |G_{wr}(e^{j\theta})|^2$ and Matlab's {\tt  psd} command, respectively.

The transfer function $G_{\text{BP}}$ in  \Fig{fig:zetamodel} is a filter designed to smooth the balancing reserve signal.   A low pass filter was adopted with crossover frequency near the nominal period of a single load: $1/(24 ~\text{hours})$ in this example.    In most of the experiments reported here, $G_{\text{BP}}$ is the first-order Butterworth low-pass filter  with cut-off frequency $f_c = 1/ (1000 ~\text{minutes})$:
\begin{equation}
	G_{\text{BP}}(z)=0.0155 \frac{1+z^{-1}}{1-0.9691z^{-1}}. 
 \label{e:GBP(z)}
\end{equation}

The reference signal must also be scaled so that the desired goal $\tilde{y}^N_t \approx r_t$ is feasible for all $t$.     Denote by $\bfmr^1$ the signal obtained using the largest scaling, while also ensuring that this signal can be tracked by the collection of pools.  This was obtained through trial and error.  

In the tracking plots, such as \Fig{fig:OptOut100_eps1}, the signals $\tilde{y}_t^N$ and $r_t$ are re-scaled to their original units in MWs.

The construction of a stationary model for the input process $\bfzeta$ was based on the linearized mean-field model,
and the scaled reference signal defined by $r_t = \epsy r^1_t$, $t\in\intgr$.    
The linear state space model \eqref{e:LSSmfg}  leads to a transfer function from  $\bfzeta$ to $\bfgamma$ that is denoted $G_p$  (recall that $\bfgamma$ is intended to approximate $\tilde{\bfmy}$).       Based on this linear approximation,  we obtain a linear mapping $\bfmr \to \bfzeta$ via the transfer function $G_c/(1+G_cG_p)$, where $G_c$ is the PI controller with proportional gain 50 and integral gain 1.5.     

\spm{Don't we need to explain PI parameters?  Is this what we have used before?  Also, not clear nor necessary:
 This linear mapping implies the variable $\epsy$ used in \textbf{A2} to scale $\bfzeta$ also applies to the regulation signal: $\bfmr^{\epsy}  = \epsy  \bfmr^1 $.
 }

\subsection{Individual QoS}
\label{s:sim_NOoptout}

The QoS of pools without opt-out control is illustrated using the histogram on the left-hand-side of \Fig{fig:BothHist} using the reference signal $\bfmr^1$. The histogram is based on samples of the  QoS function \eqref{e:Health} from each load over one month. The dashed lines represent Gaussian densities with zero mean, and variances $\sigma_{\Health}^2 $ obtained using the approximation given in \eqref{e:VarQoS}. 

%The QoS of pools without opt-out control is illustrated by histograms in \Fig{fig:Hist_p1and1} using two reference signal examples: $\bfmr^{0.3}$ and $\bfmr^1$. These two histograms are based on samples of the  QoS function \eqref{e:Health} from each load over one month. The dashed lines represent Gaussian densities with zero mean, and variances $\sigma_{\Health}^2 $ obtained using the approximation given in \eqref{e:VarQoS}. 

%\begin{figure*} 
%\begin{figure} 
%\Ebox{.85}{QoS_NoOptOut_epsy2=p09and1}  
%\vspace{-.2cm}
%\caption{Gaussian approximation of QoS histograms.}
%\label{fig:Hist_p1and1} 
%\end{figure} 
%\end{figure*} 

\begin{figure}[h]
\centering
\Ebox{.8}{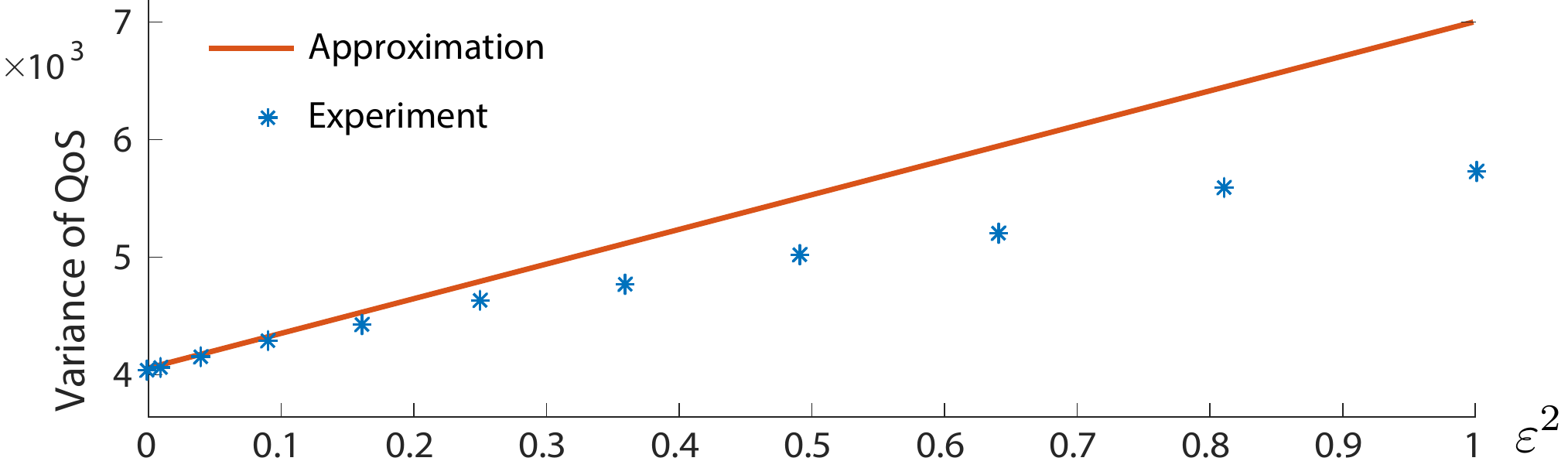}  
\vspace{-.2cm}
\caption{Linear approximation for the variance of QoS.}
%\vspace{-.1cm}
\label{fig:QoS_approx} 
\end{figure}

The conclusions of \Theorem{t:var_taylor} are illustrated in  \Fig{fig:QoS_approx}, where the QoS variances from simulation and Gaussian approximations are presented for several values of $\epsy$. It is seen that the variance is approximately linear in $\epsy^2$ for  small $\epsy>0$, with slope as predicted in \Theorem{t:var_taylor}.

\begin{figure}
\centering
\Ebox{.8}{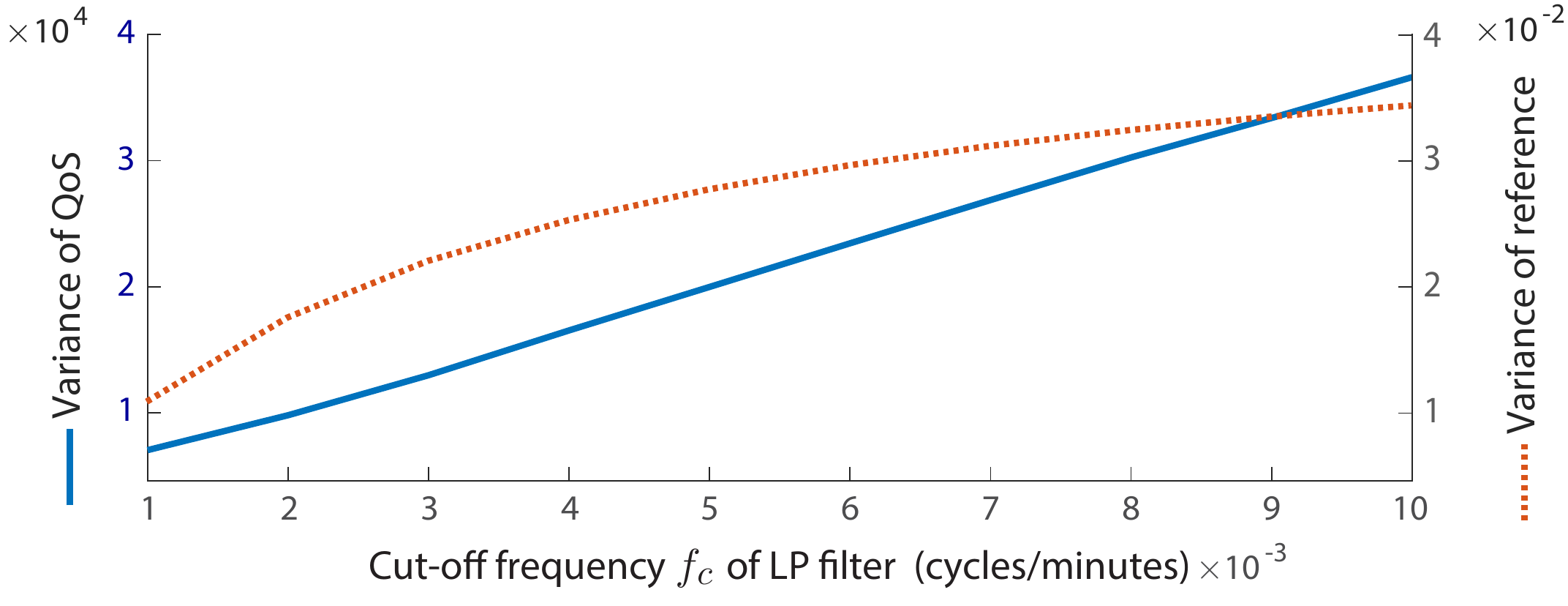}
	\vspace{-.2cm}
\caption{QoS variance increases with bandwidth of the reference signal}
	\vspace{-.1cm}
\label{fig:cut-off}
\end{figure}

\paragraph*{The role of bandpass filter}
A range of  cut-off frequencies were considered in order to investigate the impact of the bandpass filter that is used to obtain $\bfmr$.    \Fig{fig:cut-off} shows a comparison of the variance of the reference and the variance of QoS 
as a function of the cut-off frequency $f_c$ over a range of frequencies;  The linear growth in QoS variance compared to the slow growth of the variance of the reference signal justifies a filter with $f_c<10^{-2}$.

% over the range $1/ (1000 ~\text{minutes})\le f_c\le 1/ (100 ~\text{minutes})$.   
%
%The variance of the reference signal is strictly concave,  while the estimates of the QoS variance grow approximately linearly with $f_c$;  the explanation for this qualitative behavior is not known -- revise
  
%In particular, increasing the cut-off frequency to $f_c=3/ (1000 ~\text{minutes})$ will double the variance of the reference signal with a similar increase in the variance of QoS.    The value $f_c=10/ (1000 ~\text{minutes})$ results in 
%a 3-fold increase in the variance  of the reference signal,  and  a 5-fold increase in the variance of QoS.    
% 

The remaining numerical results that follow are based on the low pass filter   \eqref{e:GBP(z)} based on $f_c = 1/ (1000 ~\text{minutes})$.

\subsection{Grid level performance and opt-out control}

We present in this section experimental results with  opt-out control to ensure that QoS is subject to strict constraints.  Four QoS intervals were considered corresponding to error tolerances of, respectively, $5\%$, $10\%$, $15\%$, and $20\%$. Recall that the discount factor $\beta$ was chosen by approximating the moving window of $10$~days, which corresponds to 2880 time samples.  Based on this interpretation, these percentages are converted to intervals for local opt-out control
(following the notation  in \Section{s:optout})   as   $[b_-, b_+] =[-36, \; +36]$, $[-72, \; +72]$, $[-108, \; +108]$, and $[-144, \; +144]$, respectively. 
%\begin{figure}
%\vspace{0cm}
%\Ebox{.85}{3_LC_refx1_Hist.pdf}
%\vspace{-.2cm}
%\caption{Histogram of QoS with local opt-out control.}
%\label{fig:LC_r1_hist}
%\vspace{0cm}
%\end{figure} 

\begin{figure}[h]
\centering
 \Ebox{.75}{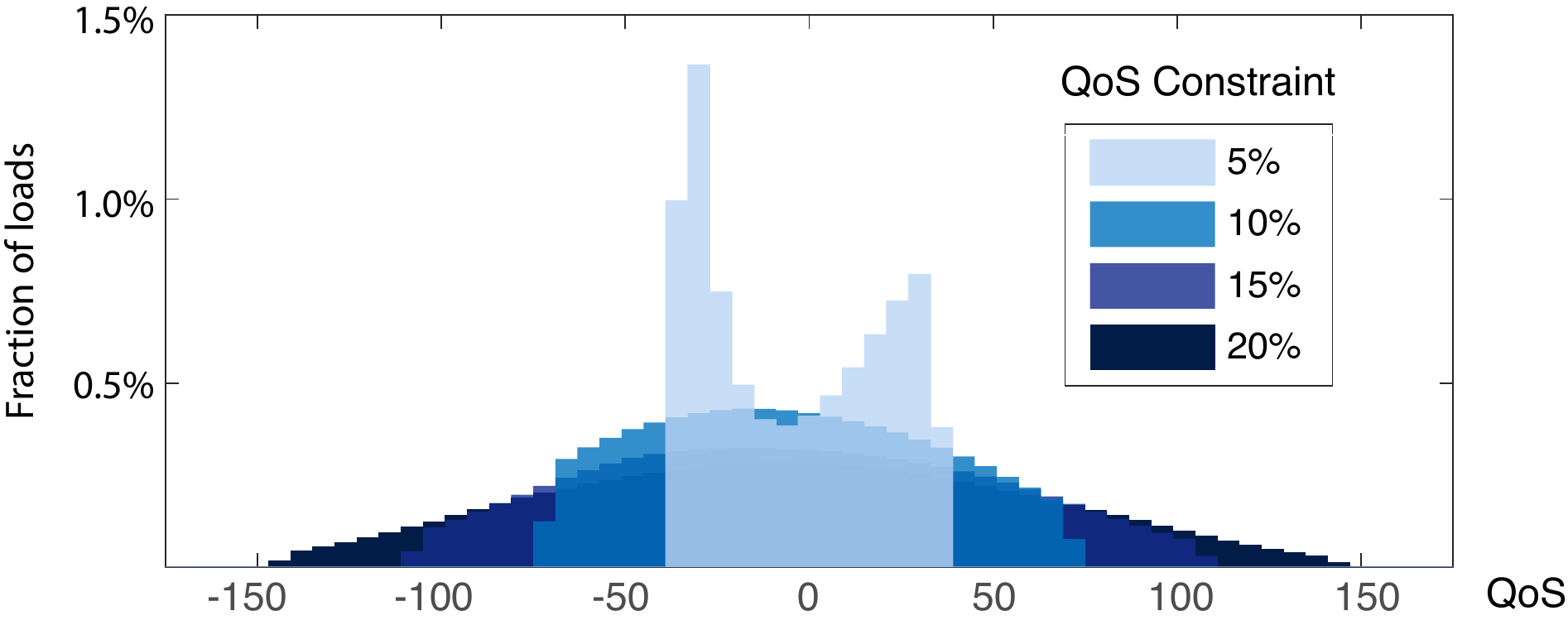}
 \vspace{-.2cm}
 \caption{QoS histograms with opt-out control using different QoS bounds.}
\label{fig:OptOut_QoS_eps1}
\end{figure}

\Fig{fig:OptOut_QoS_eps1} shows four histograms of QoS with reference signal $\bfmr^{1}$. These histograms are truncated to the predefined QoS intervals as desired.  %\Fig{fig:10daysHistBoth} provides comparison histograms of the moving-window QoS metric \eqref{e:HealthFinite0}, with and without local control.   All these histograms have mean close to 120 hours, which is the desired operation time in 10 days. The histogram without opt-out control is exactly the same as shown in \Fig{fig:10days_hist}.  The variance can be reduced by more than a factor of ten using local control, with negligible impact on grid level tracking.

%\begin{figure} 
%\Ebox{.85}{10daysHistAll}  
%\vspace{-.2cm}
%\caption{Comparison of the histograms for the \textit{moving-window} QoS indicator \eqref{e:HealthFinite0}, with and without local control. }
%\label{fig:10daysHistBoth} 
%\end{figure} 

A normalized root mean square error (NRMSE) was adopted as the metric of grid level tracking performance to study the impact of local opt-out control,
\begin{equation}
	\text{NRMSE} = \frac{\text{RMS}(\bfme) - \text{RMS}(\bfme^\bullet)}{\text{RMS}(\bfmr)},
\label{e:NRMSE}
\end{equation}
%$RMS(*) = \sqrt{\frac{1}{T}\sum_{t=1}^{T}(*_t)^2}$, 
where $\bfme^\bullet$ denotes the tracking error sequence obtained without opt-out control, and $\epsy=0$ (so that $\bfmr^\epsy \equiv 0$). 
It is found in simulation $\text{RMS}(\bfme^\bullet) \approx 8.35$ kW.

\begin{figure}[h]
\Ebox{.7}{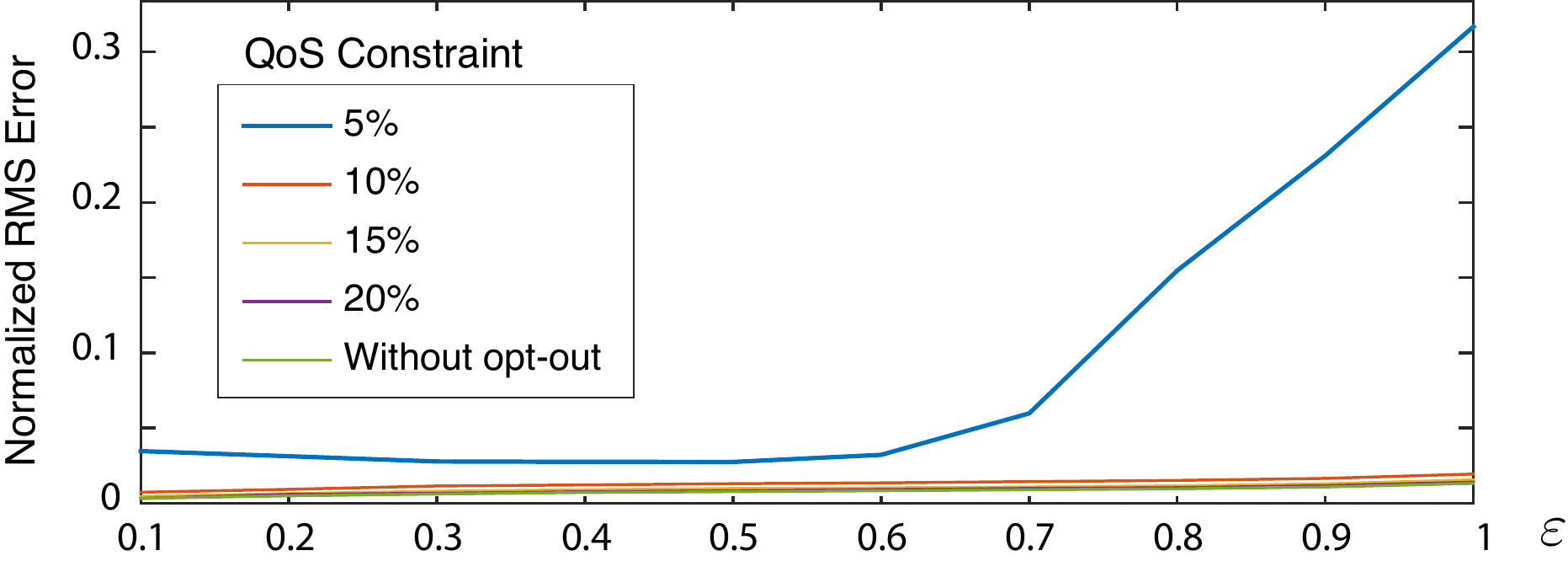}  
\vspace{-.2cm}
\caption{Tracking NRMSEs \eqref{e:NRMSE}, with and without local opt-out control.}
\label{fig:NRMSE} 
\vspace{-.2cm}
\end{figure}

The grid level tracking performance with and without opt-out control is illustrated in \Fig{fig:NRMSE}. The tracking performance with  $10\%$, $15\%$, or $20\%$ QoS interval remains nearly perfect. This is surprising, given the  improvement  in QoS shown in \Fig{fig:OptOut_QoS_eps1}. The explanation is that very few loads opt out: For example, in the simulation with $10\%$ QoS interval and reference signal $\bfmr^1$, no more than $1\%$ of loads opt out at any time. 

\notes{$\epsy^2 \to \epsy$ in x-axis of \Fig{fig:NRMSE} }

However, there are limitations on the capacity of ancillary service from  a collection of loads, and experiments reveal that opt-out control can reduce capacity.  As seen from \Fig{fig:NRMSE}, the additional opt-out control with $5\%$ QoS interval degrades the grid level tracking performance, especially when the reference signal is large. %\Fig{fig: Track_optout_epsy1} shows large tracking errors are found at time${}\approx450$~hr.

We next  apply \Proposition{t:H_approx} to better understand the grid-level impact of opt-out control, and to design an algorithm to reshape the reference signal so that loads are less likely to opt out of service.

\subsection{Re-shaping the reference input}
\label{s:num_SOC} 

Recall the SOC heuristic introduced following  \Proposition{t:H_approx}.  The proposition implies that  $\bar{\Health_t} \approx R_t$ for all $t$ when there is near perfect tracking. Conversely, this result suggests that many loads will opt-out, leading to poor tracking, if the SOC $R_t$ is near the boundary of the QoS interval $[b_-,b_+]$.

\begin{figure}[h]
\Ebox{.7}{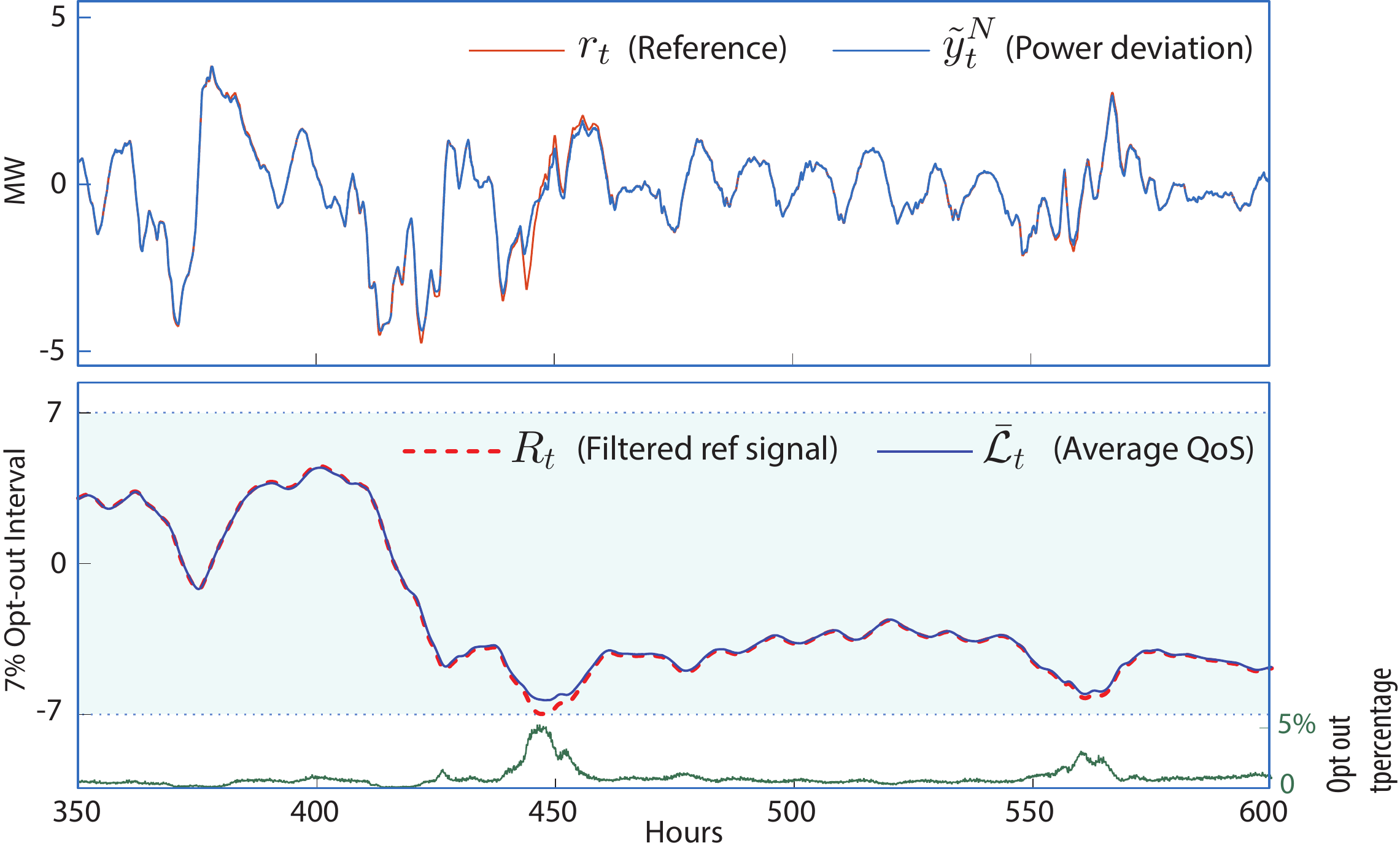}
\vspace{-.2cm}
\caption{Poor tracking results when the SOC $R_t$ reaches its  boundaries.} 
\label{fig:OptOut100_eps1}
\end{figure}

To illustrate the application of these concepts,  
consider the case with $7\%$ QoS interval and reference signal $\bfmr^1$. Results are provided in  \Fig{fig:OptOut100_eps1}.
%\Fig{fig: Track_optout_epsy1} and
Most of the time, the tracking results are nearly perfect and hence the average QoS approximates the filtered reference signal,  $\bar{\Health_t} \approx R_t$.    However, at time${}\approx450$~hr., $R_t$ falls close to $-50.4$, which is the lower QoS bound for an individual load. During this time period, many loads opted out, which resulted in  degraded tracking.
%greatly  (not so greatly!)
 % that is seen in \Fig{fig: Track_optout_epsy1}. 
 %This also explains the corresponding QoS histogram in \Fig{fig:OptOut_QoS_eps1} (c), where a greater fraction of pools close to the boundary $- 100$. 

%\begin{figure}
%	\Ebox{.85}{Tracking_OptOut100p8_epsy2=1} 
%	\vspace{-.2cm}
%	\caption{Tracking result with the reference signal scale $\bfmr^1$ and $7\%$ interval local control --- Time period corresponds to the period of poor average QoS seen in \Fig{fig:OptOut100_eps1}.}
%	\label{fig: Track_optout_epsy1} 
%\end{figure} 

In conclusion, to ensure good grid level tracking, the reference signal must respect any QoS constraints.  The grid operator should re-shape the reference signal to ensure that  the SOC of loads $R_t$ does not  approach its limits.  A smooth transformation is required because the reference signal for this example should not have significant high frequency content.

Here is one approach, based on two non-negative parameters:  a threshold $\tau<1$ and a gain  parameter
$\delta>0$. The following recursive algorithm is designed to increase (decrease) the reference signal $\bfmr$ when $\bfmR$ reaches its lower (upper) threshold. The re-shaped reference signal is defined as follows:
\begin{equation}
\!\!
\bar{r}_t =
\begin{cases}   
[ r_t - \delta (R_t - \tau b_+ ) ]_+  \ & R_t >  \tau b_+  ~ \text{and} ~ r_t >0  %\text{if}~ 
\\
[ r_t + \delta (R_t - \tau b_- ) ]_- &  R_t <  \tau b_-  ~ \text{and} ~ r_t <0
\\
r_t & \text{otherwise}
\end{cases}
\label{e:reshape}
\end{equation}
where,  $[b_-, ~ b_+]$ defines the QoS interval, and $[a]_+$ ($[a]_-$) denote the positive (negative) part of a real number $a$.

%The plots in \Fig{fig:OptOut100_improve_track} illustrate the tracking performance with $7\%$ QoS interval, and the transformed reference signal $\bar{\bfmr}^1$

When $\bfmr^1$ is transformed in this way using $\delta = 0.006$ and $\tau = 0.65$ it is found in numerical experiments that this small change in the reference signal results in perfect tracking, and a much smaller percentage of loads opting out -- a  complete simulation study will be included in \cite{YueChenThesis16}.

%  Compared to $\bfmr^1$ in \Fig{fig: Track_optout_epsy1}, $\bar{\bfmr}^1$ increases the reference signal $\bfmr^1$ at time${}\approx450$~hr. This small change in reference signal avoids a large number of opt outs at these times as observed in \Fig{fig:OptOut100_eps1}, and the tracking performance is nearly perfect.

%\begin{figure}
%\Ebox{.85}{Tracking_OptOut100p8_epsy2=1_improved}
%\vspace{-.2cm}
%\caption{Tracking performance with transformed reference signal.}
%\label{fig:OptOut100_improve_track}
%\end{figure} 

 \spm{To do:  fix percentage in figure, and explain!  The units on the x-axis are very confusing}

\subsection{Multiple QoS metrics}
\label{t:2QoS}

We next investigate the impact of opt-out control based on two QoS metrics: the cleaning QoS function \eqref{e:health_Npower} and the cycling QoS function \eqref{e:health_swt}. The lower bound of cycling QoS was set to $-\infty$. The opt-out priority is assigned to the cleaning QoS metric when both QoS metrics reach their bounds at the same time. Thus, the cycling QoS metric occasionally exceeds its predefined upper bound. 

\begin{figure}[h]
	\Ebox{.75}{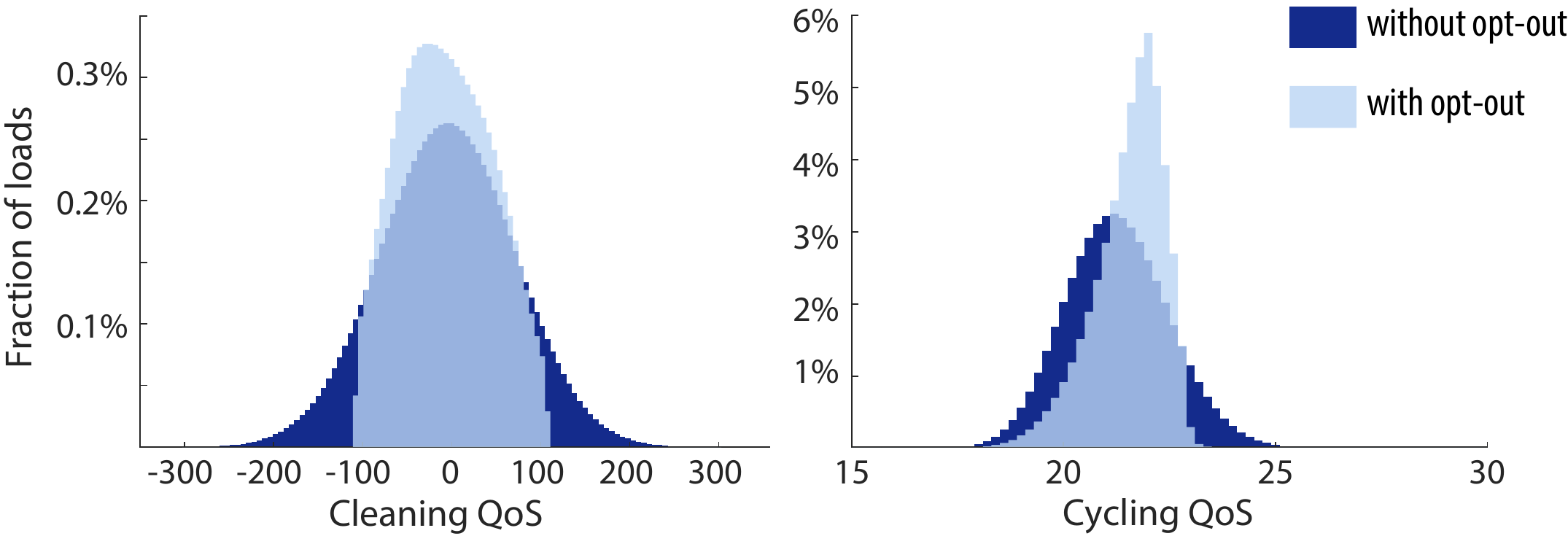}
	\vspace{-.2cm}
	\caption{QoS bounds are maintained using local opt-out control.}
	\label{fig:2QoS}
\end{figure}

Four error tolerances $\{5\%$, $10\%$, $15\%$, $20\% \}$ are considered for determining the opt-out interval $[b_-,b_+]$ for each of the two QoS metrics. \Fig{fig:2QoS} illustrates an example of QoS improvement  based on a $15\%$ error tolerance on both QoS metrics, using the reference signal $\bfmr^1$.  Tracking in this case was nearly perfect.

\begin{figure}[h]
\Ebox{.65}{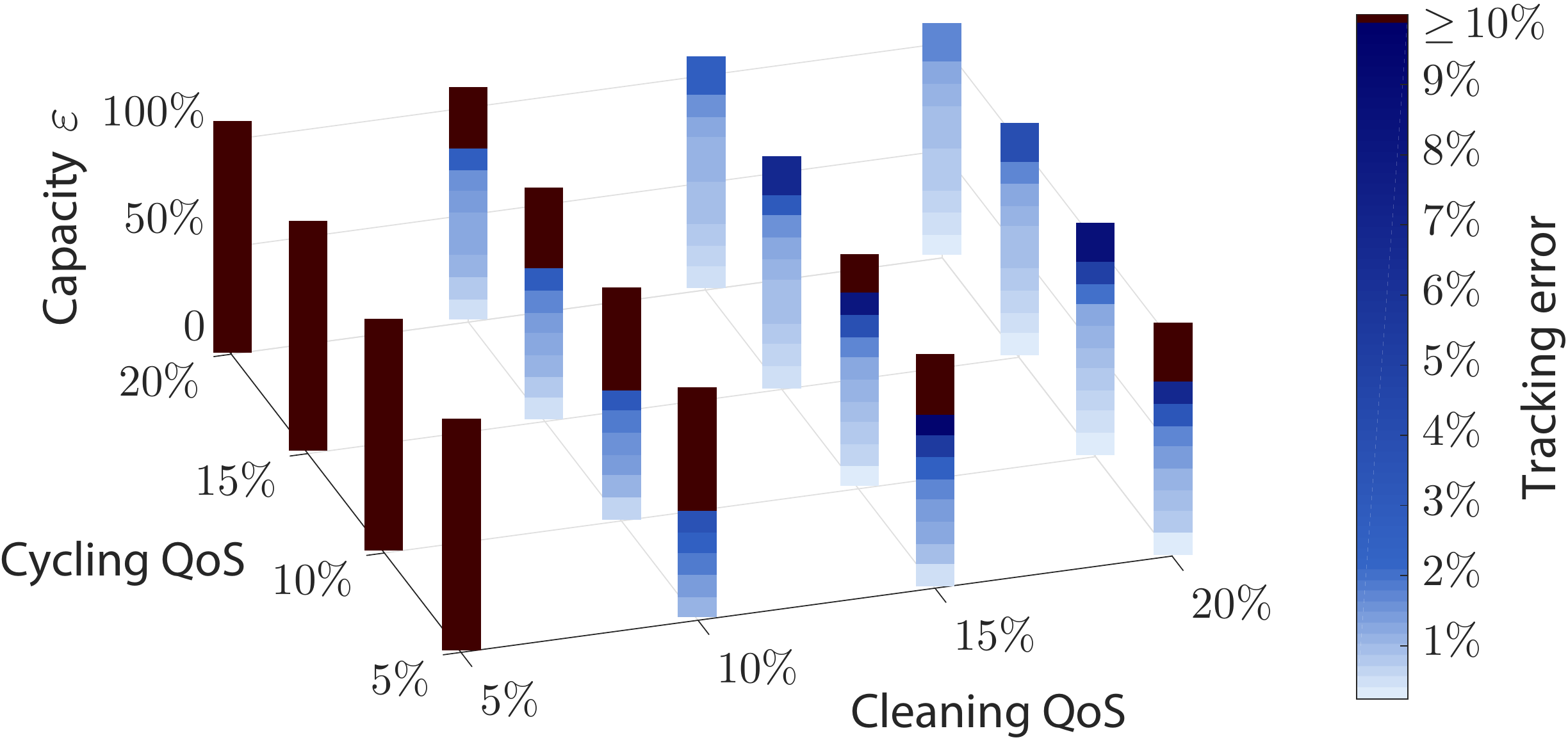} %Tracking_2QoS}
\vspace{-.2cm}
\caption{Tracking performance with two QoS constraints.}
\vspace{-.2cm}

\label{fig:track_TwoOptOuts}
\end{figure} 

Tracking performance for a range of opt-out parameters is summarized in \Fig{fig:track_TwoOptOuts} using 16 colored bars, distinguished by each pair of QoS constraints.  Each bar represents tracking errors for different reference signal scaling factors, $0.1 \le \epsy \le 1$. The darkest color represents NRMSE \eqref{e:NRMSE} of $10\%$ or greater, and lighter colors represent smaller values of NRMSE (indicated on the color bar label). Results in this figure show that opt-out control based on these two QoS metrics have little impact on tracking error over a large range of opt-out intervals. 
For those cases that local opt-out largely degrades grid-level tracking, we can either reduce the reference signal or relax QoS constraints to maintain good tracking.

%%%%%%This is caused by the large number of opt out pool in this experiment.
% the percentage of opt out pools is shown as the dashed line in \Fig{fig:LC_r2_optout}. The percentage of opt out rose to $45\%$ at $t\approx175 hr$, and kept high level until $t\approx 200 hr$. Corresponding to the large number of opt out pools at time around $175 - 200 hr$, we observed bad tracking performance at that interval in \Fig{fig: Track_optout_epsy1}.

\section{Conclusions}

The main technical contribution of this paper is the approximation of QoS for an individual load.   It is remarkable that it is possible to obtain accurate estimates of first and second order statistics for an individual load, taking  into account second order statistics of exogenous inputs (in this case the reference signal),  along with correlation introduced by the Markovian model.  It is also remarkable that strict bounds on QoS can be guaranteed while retaining nearly perfect grid-level tracking.

%Similar to batteries, loads that used to provide ancillary service can be characterized by SOC (state of charge). In our example of pools, the SOC can be estimated  in real time using the reference signal, which is available to the grid operator.  It is illustrated in simulation that the grid operator can use the SOC information to transform the original reference signal to maintain good grid-level tracking, especially when the capacity of demand dispatch is reduced by local opt-out control.

Under certain conditions, the overall QoS of a collection of loads is predictable to the grid operator. With this information, the grid operator can estimate the flexibility/capacity of loads and modify the reference signal if necessary, to maintain high quality of both grid-level tracking and QoS of loads.

Open problems remain in the area of estimation and control.   It is hoped that  estimates of the statistics of QoS can be  obtained in the presence of opt-out control. It is  also likely that control performance can be improved further with a more sophisticated approach to opt-out control.  
% {does this help the reader?
%\Fig{fig:OptOut_QoS_eps1} provides some insights: the histograms of QoS are squeezed into predefined QoS intervals, and we believe the resulting histogram shapes are also affected by the reference signal.  
%}

\medskip

\bibliographystyle{IEEEtran}
%\bibliography{strings,markov,q}

\def\cprime{$'$}\def\cprime{$'$}

\end{document}